\documentclass{article}
\begin{document}
\title{The zero point field and the reduction of the wave packet in semiclassical electrodynamics.
\author{Jacques Moret-Bailly 
\footnote{Laboratoire de physique, Université de Bourgogne, BP 47870, F-21078 Dijon cedex, France. 
email : Jacques.Moret-Bailly@u-bourgogne.fr}}}
\maketitle

\begin{abstract}
In the classical theory, an electromagnetic field obeying Maxwell's equations cannot be absorbed quickly 
by matter, so that it remains a zero point field. Splitting the total, genuine electromagnetic field into the sum 
of a conventional field and a zero point field is physically meaningless until a receiver attenuates the genuine 
field down to the zero point field, or studying the amplification of the zero point field by a source.

In classical optics all optical effects must be written using the genuine field, so that at low light levels the 
nonlinear effects become linear in relation to the conventional field. The result of the interpretation of all 
observations, even at low light levels, is exactly the same in quantum electrodynamics and in the semi-
classical theory. 

The zero point field is stochastic only far from the sources and the receivers; elsewhere, it is shaped by 
matter, it may be studied through fields visible before an absorption or after an amplification.

A classical study of the \textquotedblleft reduction of the wave packet\textquotedblright extends the 
domain of equivalence of the classical and quantum zero point field; using both interpretations of this field 
makes the results more reliable, because the traps are different.

pacs{42.25Bs, 42.50Gy}

\end{abstract}

\maketitle 

\section{Introduction.}

Many physicists think that the zero point field is introduced by quantum electrodynamics (QED); 
historically, its knowledge is anterior to QED, and the computation of its mean amplitude by Planck 
\cite{Planck} and Nernst \cite{Nernst} allowed the identification of the energy of the optical modes with 
the energy of quantum harmonic oscillators, which is the starting point of QED.

The interpretation of the zero point field is, however, fundamentally different in the classical theory and in 
QED: in the classical theory, the electromagnetic field is, in all points, a physical quantity and any solution 
of Maxwell's equations represents precisely an optical mode; in QED, the electromagnetic field is a 
mathematical tool substituted to the wave function of the photon, and the postulate \textquotedblleft 
reduction of the wave packet\textquotedblright changes the mode corresponding to a photon, thus the 
electromagnetic field. This fundamental difference is generally not clearly set, even in the best books 
(Milonni \cite{Milonni}).

For a practical use, a confusion between the quantum and classical zero point fields is not important 
because as long as the reduction of the wave packet is not used, the equations are exactly the same; this 
similitude was well studied by Marshall and Santos \cite{Marshall} and previous authors; our aim is 
seeking after an extension of their work to other optical problems, finding, in particular, a classical 
equivalent to the reduction of the wave packet, so that it becomes possible to use either interpretation in all 
problems.

The choice between the classical and QED interpretations depends on the spirit of the physicist. Using the 
interpretations of both the classical (or semi-classical) and quantum electromagnetic theories increases the 
reliability of the explanations, each theory having its specific advantages and traps; QED introduces more 
compact computations, more results use it; classical electrodynamics requires more complicated 
computations, but it seems next to the experiments; as QED is more used than the classical theory, its traps 
appear more in the literature : 

Confusing the two interpretations of the electromagnetic field, thus using wrongly the postulate 
\textquotedblleft reduction of the wave packet\textquotedblright , the best specialists of the quantum theory 
tried to discourage Townes from discovering the maser \cite{Townes}. To avoid such errors Lamb and al. 
\cite{WLamb,WLamb1} propose to reject the photon; in despite of their affirmation, is seems a {\it de 
facto} rejection of the principles of quantum electrodynamics.

For some physicists QED is useless in high field experiments: \textquotedblleft The subtle interplay 
between real and imaginary parts of the complex linear and nonlinear susceptibilities follows quite naturally 
from the semiclassical treatment. How can this information be obtained from a theory in which the fields 
are quantified ? [...] The semiclassical theory which is used in this monograph will describe all situations 
correctly in a much simpler fashion\textquotedblright (Bloembergen \cite{Bloembergen}).

\medskip
To describe correctly the semiclassical theory, we qualify the electromagnetic field as follows :

-\textquotedblleft genuine\textquotedblright   for the total electromagnetic field, with the genuine electric 
field noted $\widehat E$;

\textquotedblleft zero point\textquotedblright   for the field which remains after the largest, coherent, 
physically possible absorption of the genuine field; notation: $E_0$;

\textquotedblleft conventional\textquotedblright   for the genuine minus the zero point; as it is the commonly 
considered field, the conventional electric field will be simply written $E$;

\textquotedblleft stochastic\textquotedblright   for a stochastic zero point field.

\medskip
Section 2 reminds the origin and properties of the classical fields, in particular the zero point fields.

Section 3 gives the correct relation between the signal of a photoreceiver and the electromagnetic field.

Section 4 applies the result of section 3 to the computation of the fourth order interferences.

Section 5 proposes a classical equivalent of the postulate \textquotedblleft reduction of the wave 
packet\textquotedblright .

\section{Recall of the properties of the  classical electromagnetic field.}

As Maxwell's equations are linear in the vacuum, a linear combination of modes is a solution; with given 
limiting conditions, complete sets of modes allow to develop any mode. The energy of a mode is finite if its 
time and space extensions are limited; supposing that the mode is alone in the space, its energy is:
$W=\int {w(\nu) {\rm d}\nu}$ where $w(\nu)$ is the energy ${1\over 
2}\int(\epsilon_0E^2+\mu_0H^2){\rm d}v$ relative to the fields whose frequencies are between $\nu$ 
and $\nu +{\rm d}\nu$, at any instant. The usual normalisation is such that $\int {\frac{{w(\nu) {\rm 
d}\nu}}{\nu}}=h$. Two modes are orthogonal if the energy of their sum is the sum of their energies.

A punctual source $S_O$ of electromagnetic field (i. e. the most general multipole) in a point $O$ may be 
developed linearly using the derivatives of the three-dimensional Dirac's distribution $\delta_O$:
\begin{equation} 
S_O= \sum_{p, q, r \geq 0} f_{t, O, p, q, r} \partial_x^p\partial_y^q\partial_z^r \delta_O
\end{equation} 
where $ f_{t, O, p, q, r}$ is a distribution relative to the variable $t$, and where $ f_{t, O, 0, 0, 0}$ is a 
constant.

Here the $ f_{t, O, p, q, r}$ are supposed regular functions of time $f_{O, p, q, r}(t)$, and the number of 
non-zero $ f_{O, p, q, r}(t)$ is supposed finite. For instance, the $Oz$ oriented dipole of a molecule in 
$O$ is proportional to the distribution $ f_{O, 0, 0, 1}(t)\partial_z\delta_O$, with, for instance $ f_{O, 0, 
0, 1}(t)=\sin\omega t$.

The fields radiated by these multipoles at the point source $O$ are mathematically singular in $O$, it is an 
approximation of the physical problem discussed later. These fields are often, improperly qualified 
\textquotedblleft strictly spherical\textquotedblright although they are not invariant by a rotation around any 
axis $Ou$; the reason is that the fields on two spheres of centre $O$ have the same dependence on the 
Euler angles while it is not the case for a beam focussed in $O$ which diffracts.

Consider a particular source $S_O$, and the source $S'_O$ obtained replacing $f_{O, p, q, r}(t)$ by $-
f_{O, p, q, r}(-t)$; by a time inversion, the source $S'_O$ becomes a source $S^-_O$ which absorbs a 
\textquotedblleft strictly spherical\textquotedblright field; summing the first and the last problems, $S_O$ 
and $S^-_O$ cancel, it remains a \textquotedblleft strictly spherical\textquotedblright field $F_O$ which 
converges to $O$ then diverges.
Thus, the light emitted by a multipole will be considered as produced by the evolution of a \textquotedblleft 
strictly spherical\textquotedblright converging beam, without the multipole. This field $F_O$ is regular 
everywhere, except in $O$.

The field $F_O$ corresponding to $S_O$ is absorbed if it is completely cancelled by an opposite field; 
this opposite field cannot be a linear combination of a finite number of fields $F_A, F_B, F_C...$ 
corresponding to multipoles placed in points $ A, B, C... $ different of $O$, because these fields are 
regular in $O$. Thus {\it the absorption of the electromagnetic field radiated by a point source requires an 
infinity of point absorbers, the residual field constitutes the zero point field}. The building of the zero point 
field shows that it is an ordinary field.
\medskip

An objection to the previous mathematical description may be set:

The dipoles introduced in spectroscopy are not punctual; but their dimensions are supposed small in 
comparison with their distances, so that the conclusion remains approximately valid: the absorption of the 
field radiated by a small source requires a lot of small absorbers, a long time during which it remains a part 
of the field radiated by the dipole.

\medskip

Far from sources, the zero point field $E_0$ is generated by a large number of sources supposed 
incoherent, so that it is stochastic, it is characterised by its mean amplitude. If an atom emits a photon (that 
is, in the classical theory, a wave corresponding to an energy $h\nu$), the radiated field, large near the 
atom, is not immediately compensated by propagation, diffusion or absorption; thus the fluctuations of the 
field are shaped by next or coherent sources, so that \textquotedblleft stochastic\textquotedblright is not a 
sure property of the zero point field.

A macroscopic consequence of the structuring of the zero point field by matter is observed in the Casimir 
effect \cite{Casimir}: long wavelengths are rejected from inside two parallel plates, so that a lower 
pressure of radiation attracts the plates.

Following Einstein, the stimulated amplification of a mode by a source depends only on the amplitude of 
the true, genuine incident field $\widehat E$; the conventional, usual field $E = \widehat E - E_0$ is a 
purely mathematical object, it has no physical, individual existence.

Remark that the field radiated by a dipole is the same for an emission, a refraction or an absorption; in the 
last case the dipolar field cancels a part of an external field; the equality of Einstein's $B$ coefficients for 
stimulated emission and absorption (demonstrated by thermodynamics) appears natural. The absorption of 
light is considered now as a decrease of the energy in a mode down to the zero point energy, the emission 
is an amplification of the energy in a mode; thus all systems are connected at least by the zero point field, 
even in classical physics isolated systems do not exist.

The oscillating dipole is a particular system of moving charges; any moving electron radiates a field, but if it 
belongs to a stationary system the interference of the radiated field with the zero point field does not 
change the energy of this last field, in the average. Sommerfeld's electron does not fall on the proton, but 
the fluctuations of its interaction with the zero point field produce the Lamb shift \cite{Power,Milonni}.

Remark that the existence of the zero point field depends on the emission of an electromagnetic field by 
charged particles; if the charged particles are unable to emit a very high frequency wave, there is no zero 
point for high frequencies, no UV divergence.

Planck's constant $h$ connects the density of zero point energy in the universe to microphysics; is $h$ a 
cosmological or a microphysical constant?

\section{Absorption and detection.} 

The signal of a photoelectric cell cannot be a function of the conventional field $E$ which has no physical 
existence; for a long time, the sensitivity of the detectors of light was bad so that the zero point field could 
be neglected and the genuine field $\widehat E$ could be replaced by the conventional field

A light receiver is excited by an attenuation of the energy of a mode from a value higher than $h\nu/2$ to 
nearly $h\nu/2$. Generally this is possible if the mode was amplified by a source, but in the dark, cold, 
good photocells generate a noise signal which seems produced by the particularly large fluctuations of the 
zero point field.

The net available energy on a receiver is proportional to the difference between the input and output 
energies in the exciting mode; set $E_0$ the total zero point field received by the cell, and $E'_0$ the 
small fraction of this field amplified by the source, with an amplification coefficient $1+\gamma$; the 
available energy on the source is
\begin{equation} 
\widehat E^2-E_0^2=(\gamma E'_0+E_0)^2-E_0^2=2\gamma E'_0E_0+(\gamma E'_0)^2.
\end{equation} 
Usually $E_0$ is much smaller than $\gamma E'_0$, $2\gamma E'_0E_0$ is neglected, the usual rule is 
got; on the contrary, supposing that $\gamma$ is small, $(\gamma E_0)^2$ is neglected; for long 
observations, the time- average of the total zero point amplitude is zero, so that no signal is observed; for a 
short observation the phases are constant so that the detected signal is proportional to $\gamma E'_0$ that 
is to the amplitude , not the intensity of the conventional field, with, however, the additional factor $E_0$ 
which has an arbitrary phase. The phase factors of $E_0$ and $E'_0$ fluctuate almost independently: 
without a sophisticated detection nothing appears.

The classical emission or absorption is modelled by an excitation of a mono- or polyatomic molecule by an 
electromagnetic field up to a barrier between the two involved relative minimums of potential $u_1$ and 
$u_2$; if the initial and final states are stationary $|u_2 - u_1| = h\nu$ . But our macroscopic experiments 
usually use nearly plane modes while the molecules emit light generally through dipoles or quadrupoles. 
Quantum mechanics transforms the geometry of the waves using the \textquotedblleft reduction of the 
wave packet\textquotedblright  . Section 5 gives the classical reduction of the wave packet.

\section {Fourth order interferences.}

A sophisticated detection is performed in the fourth order interference experiments with photon counting: 
two elementary measurements are done while the phases of $E_0$ and $E'_0$ are constant (see, for 
instance, \cite{Clauser,Gosh,Ou1,Ou2,Kiess}). The result of these experiments is easily got {\it 
qualitatively} using the conventional rules \cite{M942}, but the contrast of the computed fringes is lower 
than shown by the experiments. In the simplest experiment \cite{Gosh} two small photoelectric cells are 
put in the interference fringes produced by two point sources; the interferences are not visible because they 
depend on the fast changing difference of phase $\phi$ of the modes of the zero point field amplified by the 
sources. The sources are weak; the signal is the correlation of the counts of the cells, so that the phases of 
$E_0$ and $E'_0$ are eliminated.

Distinguishing the photoelectric cells by an index j equal to 1 or 2, set $\delta_j$ the difference of paths for 
the light received by the cells. The amplitude of the conventional field received by a cell is proportional to 
$\cos (\pi\delta_j/\lambda+\phi/2)$, so that, assuming the linearity in the conventional field, the probability 
of a simultaneous detection is proportional to 
\begin{equation} 
E_0^2{E'_0}^2\cos (\frac{\pi\delta_1}{\lambda}+\frac{\phi}{2})\cos 
(\frac{\pi\delta_2}{\lambda}+\frac{\phi}{2})\propto\cos 
(\frac{\pi\delta_1}{\lambda}+\frac{\phi}{2})\cos 
(\frac{\pi\delta_2}{\lambda}+\frac{\phi}{2}).\label{interf}
\end{equation} 
The mean value of this probability got by an integration over $\phi$ is zero for $\delta_1-
\delta_2=\lambda/2$, so that the visibility has the right value 1. Assuming the usual response of the cells 
proportional to the square of the conventional field, the visibility would have the wrong value 1/2 
\cite{Mandel}.
Equation \ref{interf} may be found directly computing the interference of the zero point modes amplified 
by the sources.
\section{Classical reduction of the wave packet.}

The reduction of the wave packet breaks the symmetry of the waves, transforming, in particular, a local 
wave into a beam, for instance a spherical dipolar wave into a plane wave.

The polarisation of a transparent matter by a light beam may be observed by a variation of the energy 
levels, or detecting Kerr effect,...Thus the beginning of a pulse of light must transfer energy to matter, and 
this energy is recovered in the tail of the pulse (except for a small incoherent Rayleigh scattering). In the 
tail, the field is amplified, although there is no transition, no inversion of population, the polarisation mixing 
only slightly the initial state of the molecules with other states. This power of amplification applies not only 
to the exciting mode, so that many modes, usually initially at the zero point, are amplified, later reabsorbed 
in the medium : there is a dynamical equilibrium between the exciting field, the other modes and the 
polarisation of the molecules. The modes which are excited are dipolar or quadrupolar, they radiate far 
only the small incoherent Rayleigh field : they may be qualified \textquotedblleft local\textquotedblright  . 
On the contrary, the interactions with the light pulse are strong because they are coherent.

As the local modes are amplified, the strongest and longest fluctuations of their field may be able to excite 
molecules up to a barrier, such that an absorbing transition occurs; the mean energy of the local field, then 
of the molecules, is decreased, the amplification of the tail of the pulse is decreased, the medium has 
absorbed the light. In a laser, a similar process explains the coherent amplification by incoherently pumped 
molecules.

The emission of a field during the absorption of a quantum of energy is not instantaneous; during this 
emission and a short time after it, the probability for a strong and long fluctuation of the field is lowered, so 
that a sub-poissonian photon statistic appears; neglecting the space-time structuring of the field leads to the 
poissonian statistic \cite{Short, Glauber}. 

\section{Low level \textquotedblleft Impulsive Stimulated Raman Scattering\textquotedblright 
}

Consider a nonlinear property which may be developed as a series of the electric field of an 
electromagnetic field. It must be a function of the genuine field. The development of this function using the 
conventional field is linear for low values of this field: there is a linearisation of the properties at low values 
of the conventional field.

\medskip
If the property cannot be written as a series development, a specific treatment is necessary; consider, for 
instance the adaptation to the natural incoherent light of the \textquotedblleft Impulsive stimulated Raman 
Scattering\textquotedblright (ISRS); ISRS is known since 1968 \cite{Giordmaine} and now commonly 
used \cite{Yan,Nelson}. It is not a simple Raman scattering, but a parametric effect, combination of two 
{\it space-coherent} Raman scatterings, so that the state of the interacting molecules is not changed. The 
hot\footnote{The temperature of a spectral line is deduced from Planck's laws} exciting beam and its 
scattered beams interfere into a single frequency, redshifted, beam; the cold beam is blueshifted.

The interference of two collinear light beams into a single frequency beam is often performed involuntarily, 
for instance using a Michelson interferometer in with a moving mirror produces a Doppler frequency shift. 
The mixture of the frequencies requires an observation shorter than the beats of the two beams. The 
resulting frequency is intermediate, in proportion of the conventional amplitudes of the initial frequencies. In 
ISRS, the scattered amplitude which is permanently mixed with the incident amplitude remains extremely 
low and the Raman frequency is much lower than the exciting frequency, so that the relative frequency shift 
$\Delta\nu/\nu$ is nearly proportional to the conventional scattered amplitude and to the Raman 
frequency\footnote{Computing this interference process with the genuine amplitudes is more difficult 
because it is necessary to take into account the zero point field of the three involved modes. }.

The conventional scattered amplitude is proportional to the square of the incident genuine amplitude, 
square which, at a low level, is proportional to the conventional amplitude. More simply, while at high 
intensities the scattered amplitude is stimulated, at low intensity it is spontaneous, but it remains coherent in 
the absence of collisions because the behaviour of all molecules on a wave surface is the same (just as in 
refraction). Thus, the relative frequency shift is nearly constant, only slightly modified by the dispersion of 
the tensor of polarisability.

ISRS is obtained using ultrashort light pulses, that is \textquotedblleft pulses shorter than all relevant time 
constants\textquotedblright   \cite{GLamb}, usually femtosecond laser pulses. The relevant time constants 
in a gas may be adapted to incoherent light :

i) to avoid that the collisions destroy the coherence of the excitation of the molecules during the pulses, the 
pressure must be very low;

ii) to obtain an interference of the scattered and incident lights into a single frequency light, the period which 
corresponds to the virtual Raman transition must be larger than the length of the impulsions. The molecules 
must have transitions in the radiofrequencies, generally hyperfine transitions.

\medskip
The Universe, provides good experimental conditions for a confusion of this interaction with a Doppler 
effect: the paths are long and the pressures often low, a lot of observed polyatomic molecules or atoms 
perturbed by Zeeman effect have hyperfine structures. The absorption spectra of the molecules which are 
destroyed at their first collision, H$_2^+$ for instance, cannot be seen because the redshift simultaneous 
with the absorption of their spectra widens, thus weakens, their lines.

\section{Conclusion.}
 The electromagnetic fields, in particular the zero point field, often, improperly qualified \textquotedblleft 
stochastic\textquotedblright, obey the same equations in quantum and classical theories. Both, very 
different interpretations, are useful, giving the same final results, having their specific advantages : Quantum 
electrodynamics provides ready to use properties or postulates, but a common improper use of some of its 
concepts, the photon for instance, leads to wrong conclusions \cite{WLamb}; classical electrodynamics is 
more intuitive, but it requires often more complicated demonstrations.

An isolated system cannot exist in classical optics because the electromagnetic fields expand everywhere, 
although they are amplified or attenuated by matter.

The teachers should point the approximation made neglecting the zero point field and replace unnecessarily 
approximate rules, for instance the first Planck's law, by the rigorous rules.

\end{document}